\documentclass[prl,aps,twocolumn,showpacs,floatfix]{revtex4-1}

\usepackage{amsfonts}
\usepackage{amssymb}
\usepackage{hyperref}
\usepackage{graphicx}
\usepackage{dcolumn}
\usepackage{bm,amsmath,verbatim}
\usepackage{mathrsfs}
\usepackage{color}

\usepackage{amsmath,amssymb,amsthm}

\usepackage[T1]{fontenc}
\usepackage[latin9]{inputenc}
\usepackage{booktabs}

\begin{document}

\title{Anomalous Transport Induced by Non-Hermitian Anomalous Berry Connection in Non-Hermitian Systems}
\author{Jiong-Hao Wang$^{1}$}
\author{Yu-Liang Tao$^{1}$}
\author{Yong Xu$^{1,2}$}
\email{yongxuphy@tsinghua.edu.cn, yongxuphy@mail.tsinghua.edu.cn}
\affiliation{$^{1}$Center for Quantum Information, IIIS, Tsinghua University, Beijing 100084, People's Republic of China}
\affiliation{$^{2}$Shanghai Qi Zhi Institute, Shanghai 200030, People's Republic of China}

\begin{abstract}
Non-Hermitian materials can not only exhibit exotic energy band structures but also an anomalous velocity induced by non-Hermitian
anomalous Berry connection as predicted by the semiclassical equations of motion for Bloch electrons. However,
it is not clear how the modified semiclassical dynamics modifies transport phenomena. Here, we theoretically
demonstrate the emergence of anomalous oscillations driven by either an external dc or ac electric field, which arise from
non-Hermitian anomalous Berry connection. Moreover, it is a well-known fact that geometric structures of electric wave functions
can only affect the Hall conductivity. However, we are surprised to find a non-Hermitian anomalous Berry connection
induced anomalous linear longitudinal conductivity independent of the scattering time.
We also show the emergence of a second-order nonlinear longitudinal conductivity induced by
non-Hermitian anomalous Berry connection, violating a well-known fact of its absence in a Hermitian system with symmetric energy spectra. These anomalous phenomena are illustrated in a pseudo-Hermitian system with large non-Hermitian anomalous Berry
connection. Finally, we propose a practical scheme to realize the anomalous oscillations in an optical system.
\end{abstract}

\maketitle

Non-Hermitian physics has recently been one of active subjects intensively studied in various
branches of physics ranging from optical and acoustic systems, cold atomic systems to condensed matter materials~\cite{ChristodoulidesNPReview,XuReview,ZhuReview,UedaReview,BergholtzReview}.
The existence of non-Hermitian terms in a Hamiltonian, such as gain or loss, can lead to exotic energy band
structures that have no counterparts in Hermitian systems, such as band structures with exceptional points or
rings~\cite{Zhen2015nat,Xu2017PRL,Cerjan2018PRB,Zhou2018,Carlstrom2018PRA,
HuPRB2019,Wang2019PRB,Ozdemir2019,Cerjan2019nat,
Kawabata2019PRL,Zhang2019PRL,Chuanwei2020PRL,Yang2020PRL,Wang2021PRL}.
Recently, it has been shown that non-Hermitian physics may also arise in disordered or strongly correlated systems due
to finite lifetimes of quasiparticles~\cite{Kozii2017,Zyuzin2018PRB,Yoshida2018PRB,Zhao2018PRB,Yoshida2019PRB,
Nagai2020PRL,Sato2021PRL,Tao2021}. This motivates us to ask how transport phenomena are modified in non-Hermitian systems.

The semiclassical dynamics of Bloch electrons in external fields has proven to be a powerful
theoretical framework to account for various transport properties~\cite{Niu1995PRL,Niu1999PRB,Xiao2005PRL,Xiao2010RMP,
Gao2014PRL,Fu2015PRL}. For instance,
the semiclassical equations of motion predict an anomalous transverse velocity arising from
the geometric structures of electric wave functions~\cite{Xiao2010RMP}.
The geometric structures are involved in the semiclassical equations of motion in terms of
Berry curvature rather than Berry connection, which is gauge dependent.
Such an anomalous transverse velocity can only induce a Hall current
instead of a longitudinal current.
Given that non-Hermitian physics can exist in various systems, it is
important to ask how the semiclassical dynamics should be modified in a non-Hermitian system.

In fact, in 2017, one of the authors derived the following semiclassical equations of motion for Bloch electrons
in an external electric force $\bm {F}=-e{E}_\mu {\bm e}_\mu$ in Ref.~\cite{Xu2017PRL} (see also Refs.~\cite{XuReview,Silberstein2020PRB,Supplement}):
\begin{subequations} \label{SME1}
\begin{align}
\dot { {r}}_\lambda=&\partial_\lambda \bar{\varepsilon}_n-
e\epsilon_{\lambda\mu\nu}\Omega_{n,\mu}{E}_\nu, \label{SME1a}\\
\dot {{k}}_\lambda=&-e{E}_\lambda,
\end{align}
\end{subequations}
where $\bm {r}=r_\mu{\bm e}_\mu$ and $\bm {k}=k_\mu{\bm e}_\mu$ denote the center of mass of a wave packet
in real and momentum space, respectively, $\epsilon_{\lambda\mu\nu}$ is the Levi-Civita symbol, and ${\bm e}_\mu$ is the unit vector along the $\mu$ direction ($\mu=x,y,z$).
Here, to simplify notations, we have set $\hbar=1$, defined $\partial_\lambda=\partial_{k_\lambda}$,
adopted the Einstein summation convention and will set the lattice constant to one henceforth.
$\Omega_{n,\lambda}=i\varepsilon_{\lambda\mu\nu} \langle{\partial_\mu u_{n}^R}| {\partial_\nu u_{n}^R}\rangle$ is
the Berry curvature in the $n$th band, which accounts for the intrinsic anomalous Hall effects
(Note that
only the Berry curvature defined by the right eigenstates is relevant to the velocity~\cite{Supplement}).
Here, $|{u_{n}^R}\rangle$
is the normalized right eigenstate of a generic non-Hermitian Hamiltonian $H({\bm k})$
in momentum space in the $n$th band, i.e.,
$H({\bm k})|{u_{n}^R(\bm k)}\rangle=\varepsilon_n({\bm k}) |{u_{n}^R(\bm k)}\rangle$ with
$\langle{u_{n}^R(\bm k)}|u_{n}^R(\bm k)\rangle=1$. In fact, for a non-Hermitian
Hamiltonian, there appears a normalized left eigenstate $\langle{ {u}_{n}^L(\bm k)}|$ satisfying
$\langle{ {u}_{n}^L(\bm k)}|H(\bm {k})=\langle{ {u}_{n}^L(\bm k)}| \varepsilon_n({\bm k})$ and
$\langle {u}_n^L(\bm {k})|u_n^R(\bm {k})\rangle=1$, which coincides with the Hermitian
conjugate of the corresponding right eigenstate in the Hermitian case.
The emergence of the different left eignstate leads to an effective energy spectra
(note that the second part does not contribute to the distribution function)
\begin{equation}
\bar{\varepsilon}_n({\bm {k}})=\text{Re}[\varepsilon_n({\bm {k}})]
-e\bar{%
{A}}_{n,\mu}(\bm {k}){E}_\mu,
\end{equation}
where a non-Hermitian anomalous Berry connection (NHABC) arises
\begin{equation}
\bar{{A}}_{n,\mu}(\bm {k})\equiv\text{Re}[{A}_{n,\mu}(\bm {k})-\tilde{{A}}_{n,\mu}(\bm {k})].
\end{equation}
Clearly, the Berry connection is involved in the equation through the difference of the
right-right Berry connection ${A}_{n,\mu}(\bm {k})=i\langle{u}_n^R(\bm {k})|\partial_\mu u_n^R(\bm {k})\rangle$
and the left-right Berry connection
$\tilde{{A}}_n(\bm {k})=i\langle{u}_n^L(\bm {k})|\partial_{\mu}u_n^R(\bm {k})\rangle$,
showing the fact that this term can only appear in non-Hermitian systems.
Such a term is nonzero in a generic non-Hermitian system [except in a $\mathcal{PT}$ (product of inversion and
time-reversal symmetry) or $C_2\mathcal{T}$ (product of two-fold rotational and time-reversal symmetry) symmetric system]~\cite{Supplement}.
Based on Eq.~(\ref{SME1a}), this term leads to a non-Hermitian anomalous velocity
\begin{equation}
{v}_{\mathrm{NA},\lambda}=-e{E}_\mu \partial_\lambda \bar{{A}}_{n,\mu}.
\end{equation}

Despite the fact that the semiclassical equations of motion have been derived in Refs.~\cite{Xu2017PRL,Silberstein2020PRB},
it remains an important open question of whether such a modified dynamics will result in anomalous transport.
In the paper, we study two classes of transport phenomena: coherent dynamics of one electron, and linear and nonlinear
conductivities of many electrons.
We find that the existence of the non-Hermitian anomalous velocity results in anomalous features in oscillations
driven by either dc or ac electric fields. For the linear longitudinal conductivity,
we are surprised to find a NHABC induced anomalous longitudinal conductivity
that is independent of the scattering time.
In addition, it is a well-known fact that in a Hermitian system with symmetric energy spectra, a second-order
nonlinear longitudinal conductivity is forced to vanish. Remarkably, we discover
a second-order nonlinear longitudinal conductivity induced by the NHABC.
These results suggest that the geometric structures of wave functions can not only induce a Hall current
but also a longitudinal current in a non-Hermitian system.
We demonstrate these anomalous phenomena in a pseudo-Hermitian system with large NHABC.

\begin{figure}[t]
\includegraphics[width=3.4in]{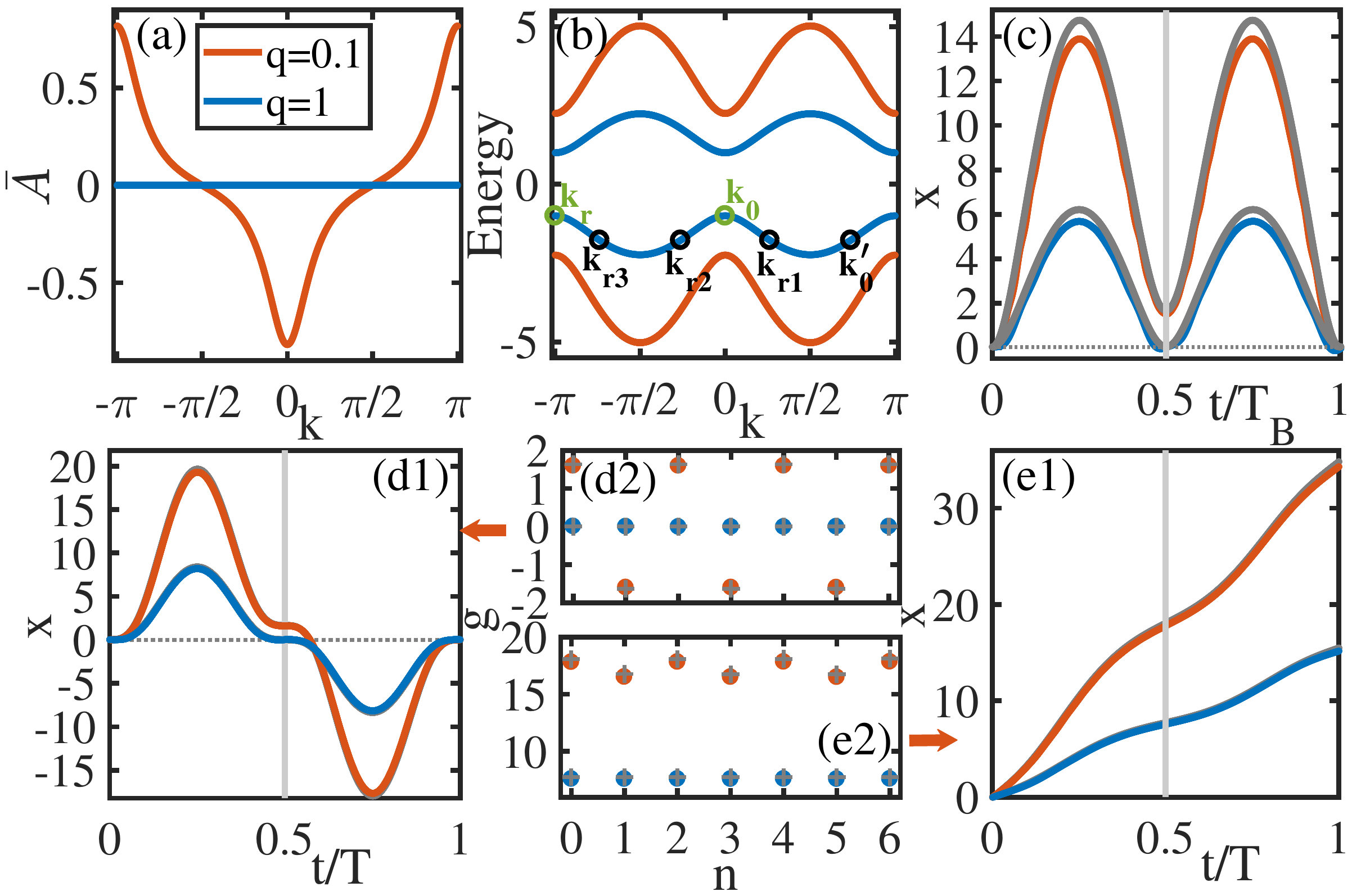}
\caption{(Color online) (a) The computed NHABC and (b)
energy spectra versus $k$. In (b), the circles denote the states with the same energy in a certain band. The time evolution of the center of mass of a wave packet in real space
(c) under a dc electric field and (d1-e1) under a sinusoidal ac electric field
with the corresponding $g$ function plotted in (d2) and (e2), respectively.
The electric field is ${E}=0.2$.
In (c), $k(0)=0$. In (d1) and (d2), $k(0)=0$ and $k(T/2)=-\pi$.
In (e1) and (e2), $k(0)=-0.2$ and $k(T/2)=-1.3$.
The blue and red lines and solid circles show the results
obtained by numerically solving the Schr\"odinger equation, while the grey ones show
the results obtained by numerically solving the semiclassical equations of motion (\ref{SME1}).
The results imply that the semiclassical dynamics agrees well with the dynamics of a wave packet governed directly by
the Schr\"odinger equation.
The blue (red) lines or solid circles correspond to the results for the
Hamiltonian (\ref{HamPse}) with parameters (\ref{dparas}) and $q=1$ (Hermitian) [$q=0.1$ (non-Hermitian)], respectively.
Here $t_0=1$, $t_1=0$ and $t_2=2$.}
\label{fig1}
\end{figure}

\emph{Model}---To demonstrate the anomalous transport properties, we start by studying the NHABC in a one-dimensional (1D) two-band
non-Hermitian system described by the following Hamiltonian in momentum space,
\begin{equation} \label{HamPse}
H(k)={\bm d}\cdot{\bm {\bar\sigma}}+d_0,
\end{equation}
where
\begin{equation}
\bar{\sigma}_x=\left(
                 \begin{array}{cc}
                   0 & a \\
                   b & 0 \\
                 \end{array}
               \right),
\bar{\sigma}_y=\left(
                 \begin{array}{cc}
                   0 & -ia \\
                   ib & 0 \\
                 \end{array}
               \right),
\bar{\sigma}_z=\left(
                 \begin{array}{cc}
                   q^{-1} & 0 \\
                   0 & -q \\
                 \end{array}
               \right),
\end{equation}
are $q-$deformed Pauli matrices with $a=\sqrt{(1+q^2)/2}$, $b=\sqrt{(1+q^{-2})/2}$ and $q>0$~\cite{Blohmann2003,Shiliang2021}.
Note that such matrices have also been used to construct non-Hermitian Chern insulators, Weyl semimetals and chiral
topological insulators~\cite{Shiliang2021}.
Let us first consider a system with time-reversal symmetry with
\begin{equation} \label{dparas}
d_x=t_0+t_1\cos k, d_y=t_2\sin k, d_z=m, d_0=0,
\end{equation}
where $t_0$, $t_1$, $t_2$ and $m$ are real parameters.
Although the Hamiltonian is non-Hermitian when $q\neq 1$, it is pseudo-Hermitian~\cite{Mostafazadeh2002}, and its energies $\varepsilon_{\pm}$ are real~\cite{Supplement}.
Such a real energy spectrum indicates the absence of the skin effects due to the absence of the winding number even though the Hamiltonian
in real space has asymmetric hopping~\cite{ChenFang2020PRL,Okuma2020PRL,Slager2020PRL}.
Although the system is topologically equivalent to its Hermitian counterpart,
we find that the NHABC emerges, that is,
\begin{equation}
\bar{A}_\pm(k)=\frac{b(a-b)(t_2 d_x\cos k+t_1 d_y \sin k)(\xi_\pm+mc_1)}{2\xi_\pm[(\xi_\pm+mc_1)^2+b^2(d_x^2+d_y^2)]},
\end{equation}
where $\xi_{\pm}=\pm \sqrt{ab(d_x^2+d_y^2)+m^2c_1^2}$ and $c_1=(1+q^2)/(2q)$.
For simplicity, consider $m=0$ so that the expression reduces to
\begin{equation}
\bar{A}_\pm(k)=c_2\frac{t_2(t_1+t_0\cos k)}{(t_0+t_1\cos k)^2+t_2^2 \sin^2 k},
\end{equation}
where $c_2=-(1-q)/(2(1+q))$ ($|c_2|<= 1/2$).
In this case, $\bar{A}_{\pm}$ does not depend on the band index, and we thus drop the band index henceforth. Clearly, when $q=1$,
the Hamiltonian is Hermitian so that the
term vanishes. Specifically, consider the NHABC at $k=0$ or $\pi$, which reads
$\bar{A}(0)=c_2 t_2/(t_0+t_1)$
and $\bar{A}(\pi)=-c_2 t_2/(t_0-t_1)$. Remarkably, one of the terms diverges when either $t_0+t_1=0$
or $t_0-t_1=0$.
When $|t_0/t_1|<1$, we find that $\bar{A}_{\pm}=c_2t_1/t_2$ at $k=\text{arccos}(-t_0/t_1)$,
which diverges at $t_2=0$. We therefore conclude that the NHABC
can be large in the model. We remark that when $\bar{A}$ diverges, the energy gap also closes
at the corresponding momentum. With an energy gap, $\bar{A}$ is distributed around the momenta associated with
minima of direct gaps, as shown in Fig.~\ref{fig1}(a).

\emph{Anomalous oscillations in an external dc electric field}---We now study
the dynamics of a wave packet in a periodic potential subject to a dc electric field so that
the wave packet undergoes a Bloch oscillation.
Without loss of generality, we consider a 1D case.
Based on the semiclassical equations of motion~(\ref{SME1}), the position and quasimomentum of a wave packet evolves as
\begin{subequations}
\begin{align} \label{SME}
x(t)=&-\frac{1}{e{E}}[\bar{\varepsilon}(k(t))-\bar{\varepsilon}(k_0)]=x_{\mathrm{H}}(t)+x_{\mathrm{NA}}(t), \\
{k}(t)=&-e{E} t+{k}_0,
\end{align}
\end{subequations}
where we have set $x(0)=0$, and $k_0$ is the initial quasimomentum of the wave packet. The position
is determined by two parts: the weighted decrease in the real part of the energy spectrum,
$x_{\mathrm{H}}(t)=-\text{Re}[\varepsilon({{k(t)}})-\varepsilon({{k_0}})]/(e{E})$,
and the increase in the NHABC,
$x_{\mathrm{NA}}(t)=\bar{{A}}({k}(t))-\bar{{A}}({k_0})$. In a Hermitian system,
$x_{\mathrm{NA}}$ vanishes so that the position is entirely determined by $x_{\mathrm{H}}(t)$ which exhibits
an oscillation with a period of $T_{\mathrm{B}}=2\pi/(e{E})$.

Another feature in the Hermitian case is that
besides at the time of integer multiples of the period at which the wave packet returns to the
initial position, this also occurs at other return times $t_r$ with
$\text{Re}[\varepsilon({{k(t_r)}})]=\text{Re}[\varepsilon({{k_0}})]$.
For example, consider the energy spectrum in
Fig.~\ref{fig1}(b). The energies at $k_0$ and $k_r$ are equal so that the wave packet has to move back to
the original position at time $t_r$ when $k(t_r)=k_r$. Similarly, the energies at $k_0^\prime$, $k_{r1}$,
$k_{r2}$ and $k_{r3}$ are equal, leading to the fact that the return happens at times $t_{r1}$, $t_{r2}$ and $t_{r3}$ corresponding to $k(t_{r1})=k_{r1}$,
$k(t_{r2})=k_{r2}$ and $k(t_{r3})=k_{r3}$.

However, for a non-Hermtian system, the rule is generically violated due to the contribution
of $x_{\mathrm{NA}}(t)$, in the sense that even though $x_{\mathrm{H}}$ vanishes, $\bar{{A}}({k_r})$ is not necessarily equal to
$\bar{{A}}({k_0})$. In fact, the return times are shifted or lifted depending on $x(t)=0$.
Indeed, Fig.~\ref{fig1}(c) illustrates that at time $t_r=T_{\mathrm{B}}/2$, $x(t_r)$ is lifted so that
$x(t)>0$ when $t\in(0,T_{\mathrm{B}})$, in stark contrast to the Hermitian case shown by the blue line.

\emph{Anomalous oscillations in an external ac electric field}---To exhibit anomalous oscillations in an ac electric field, we
require that the field should be either positive or negative in each
half cycle and is antisymmetric with respect to $T/2$ ($T$ is the time period), i.e., $E(t)=-E(T-t)$. The requirements are naturally satisfied by
commonly used ac electric fields, such as sinusoidal, triangular and square waveforms.
With the electric field, the quasimomentum of a wave packet center first moves from $k_0=k(0)$ to $k_m=k(T/2)$ in
the first half cycle and then returns to $k_0$ in the second half cycle.
One can also prove that
the displacements of the wave packet over the first and second half cycles in a Hermitian system are equal, i.e.,
$x_H(T/2)-x_H(0)=x_H(T)-x_H(T/2)$ based on the result $k(t)=k(T-t)$ for $0\le t\le T/2$. Thus, we
can define the displacement as a discrete function
\begin{equation}
g(n)=x[(n+1)T/2]-x(nT/2),
\end{equation}
which is a constant function in the Hermitian case. For example, for a dynamics of a wave packet in a Hermitian system shown
by the blue lines in Fig.~\ref{fig1}(d1-e1), the associated $g$ functions shown in Fig.~\ref{fig1}(d2-e2) are constant functions.
In a specific case with $\text{Re}[\varepsilon(k_0)]=\text{Re}[\varepsilon(k_m)$],
$g(n)=0$, showing that a wave packet returns to the initial position over each half cycle as shown in Fig.~\ref{fig1}(d2).
However, in the non-Hermitian case,
\begin{equation}\label{fEn}
g(n)=C+(-1)^n [\bar{{A}}({k_m})-\bar{{A}}({k_0})],
\end{equation}
where $C=x_{\mathrm{H}}[(n+1)T/2]-x_{\mathrm{H}}(nT/2)$ which is constant contributed by the energy dispersion.
$C$ vanishes when $\text{Re}[\varepsilon(k_0)]=\text{Re}[\varepsilon(k_m)$].
It is clear to see that $g$ is no longer a constant function when $\bar{{A}}({k_m})\neq \bar{{A}}({k_0})$, and
it varies with respect to $n$ with a period of $2$. The period change can be clearly seen in the $g$ functions (red solid circles) shown
in Fig.~\ref{fig1}(d2-e2) for a dynamics of a wave packet in a non-Hermitian system shown
by the red lines in Fig.~\ref{fig1}(d1-e1),
Such a period change of the function can be directly measured in experiments.

\emph{Anomalous linear and nonlinear longitudinal conductivities}---To investigate the electric response to an electric
field for many electrons in a system with impurities, we employ the semiclassical equations
of motion together with the Boltzmann equation. In the relaxation time approximation, the Boltzmann equation for the distribution
function $f$ of electrons is
\begin{equation}
-e\tau E_\mu \partial_\mu f+\tau \partial_t f=f_0-f,
\end{equation}
where $f_0$ is the equilibrium distribution function without external fields, and $\tau$ is the scattering time.
In a non-Hermitian case, $f_0(\varepsilon({\bf k}))=1/[\exp [(\text{Re}(\varepsilon({\bf k}))-\mu)/(k_B T)]+1]$
corresponds to the Fermi-Dirac distribution for the real part of the eigenenergy with $\mu$ being the chemical potential and $T$ being
the temperature; the imaginary part of the eigenenergy plays the role of the scattering time.
For generality, we consider an ac electric field, $E_\mu=\text{Re}\{\mathcal{E}_{\mu} e^{i\omega t}\}$ with
$\omega$ being the angular frequency. To see the effects of the NHABC,
we expand the distribution function up to the second order: $f=\text{Re}\{f_0+f_1+f_2 \}$
with $f_1=g_{1}^{\omega}e^{i\omega t}$ and $f_2=g_2^{0} +g_2^{2\omega} e^{i2\omega t}$. Based on the
Boltzmann equation above, one obtains~\cite{Fu2015PRL}
\begin{eqnarray}
g_1^\omega &=&\frac{e\tau \mathcal{E}_\mu \partial_\mu f_0}{1+i\omega \tau},
g_2^0=\frac{(e\tau)^2 \mathcal{E}_\mu^* \mathcal{E}_\nu \partial_{\mu\nu}f_0}{2(1+i\omega \tau)}, \nonumber \\
g_2^{2\omega}&=&\frac{(e\tau)^2 \mathcal{E}_\mu \mathcal{E}_\nu \partial_{\mu \nu}f_0}{2(1+2i\omega\tau)(1+i\omega\tau)}.
\end{eqnarray}
Combined with the semiclassical equation, we derive the current $j_\lambda=-e\int_k fv_\lambda=\text{Re}\{j_\lambda^0+j_\lambda^{\omega}e^{i\omega t}+j_\lambda^{2\omega} e^{i2\omega t}\}$ where
\begin{subequations} \label{JEqn}
\begin{align}
j_\lambda^0 &={e}\int_k (-g_2^0 \partial_\lambda \text{Re}(\varepsilon)+\frac{e}{2} g_1^\omega \bar{\Omega}_{\lambda\nu} \mathcal{E}_\nu^*), \\
j_\lambda^{\omega} &= e\int_k (-g_1^{\omega} \partial_\lambda \text{Re}(\varepsilon) +
ef_0\bar{\Omega}_{\lambda\nu}\mathcal{E}_\nu), \\
j_\lambda^{2\omega} &= {e}\int_k (-g_2^{2\omega}\partial_\lambda \text{Re}(\varepsilon)+\frac{e}{2} g_1^{\omega}\bar{\Omega}_{\lambda\nu}\mathcal{E}_\nu),
\end{align}
\end{subequations}
where the terms $j_\lambda^0$, $j_\lambda^{\omega}$
and $j_\lambda^{2\omega}$ describe the rectified, first harmonic and second harmonic currents, respectively, and $\int_k\equiv
\int_{\mathrm{BZ}} d^d k/(2\pi)^d$, an integral over the first Brillouin zone with $d$ being the dimension of a system.
Here we have introduced a new quantity,
\begin{equation}
\bar{\Omega}_{\lambda\nu}=\varepsilon_{\lambda\mu\nu}\Omega_\mu+\partial_\lambda\bar{A}_\nu,
\end{equation}
where the first term is associated with a Berry curvature dipole resulting in nonlinear Hall effects~\cite{Fu2015PRL},
and the second term results from the NHABC.
When $\lambda=\nu$, the quantity is completely determined by the NHABC,
i.e., $\bar{\Omega}_{\nu\nu}=\partial_\nu\bar{A}_\nu$.
For simplicity, we utilize the constant relaxation time approximation. Consider a system with $\text{Re}[\varepsilon(-{\bm k})]=\text{Re}[\varepsilon({\bm k})]$,
such as a system with either time-reversal symmetry or inversion symmetry. Then,
$\partial_{\mu\nu}f_0 \partial_\lambda \text{Re}(\varepsilon)$ is an odd function
with respect to $\bm k$, forcing the corresponding integrals to vanish. We hence obtain
$j_\lambda^0=\chi_{\lambda\mu\nu}\mathcal{E}_\mu\mathcal{E}_\nu^*$,
$j_\lambda^{\omega}=\sigma_{\lambda\mu}\mathcal{E}_\mu$,
$j_\lambda^{2\omega}=\chi_{\lambda\mu\nu}\mathcal{E}_\mu\mathcal{E}_\nu$,
where $\sigma_{\lambda\mu}$ is the linear conductivity tensor with
\begin{equation}
\sigma_{\lambda\mu}=e\int_k (-\frac{e\tau \partial_\mu f_0}{1+i\omega \tau} \partial_\lambda \text{Re}(\varepsilon) +
ef_0\bar{\Omega}_{\lambda\mu}),
\end{equation}
and $\chi_{\lambda\mu\nu}$ is the second-order nonlinear conductivity tensor with
\begin{equation}
\chi_{\lambda\mu\nu}=\frac{e^3\tau}{2(1+i\omega \tau)}\int_k \partial_\mu f_0 \bar{\Omega}_{\lambda\nu}
=-\frac{e^3\tau}{ 2(1+i\omega \tau) }\int_k f_0 \bar{D}_{\lambda\mu\nu},
\end{equation}
where $\bar{D}_{\lambda\mu\nu}=\partial_\mu\bar{\Omega}_{\lambda\nu}= \epsilon_{\lambda \alpha \nu}\partial_\mu \Omega_\alpha+\partial_{\lambda\mu}\bar{A}_\nu$ with the first term being the local Berry curvature dipole~\cite{Fu2015PRL} and the
second term induced by the NHABC.

\begin{figure}[t]
\includegraphics[width=3.4in]{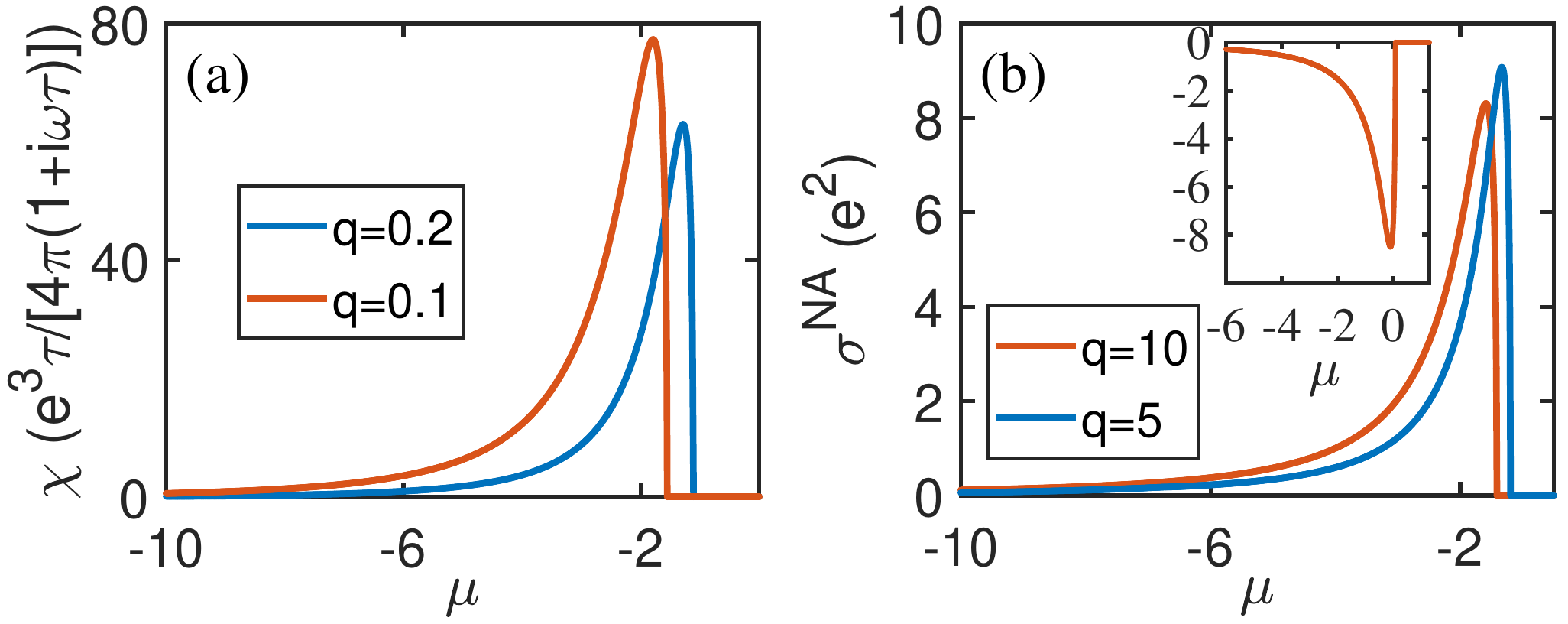}
\caption{(Color online)
(a) The second-order nonlinear longitudinal conductivity $\chi$ versus the chemical potential $\mu$ for the 1D Hamiltonian (\ref{HamPse}) in terms of parameters in (\ref{dparas})
with time-reversal symmetry at zero temperature. Here,
$t_0=1$, $t_1=6$ and $t_2=0.7$.
(b) The linear conductivity $\sigma^{\mathrm{NA}}=\sigma_{xx}^{\mathrm{NA}}$ resulted from the NHABC versus the chemical potential $\mu$ for the 1D Hamiltonian
(\ref{HamPse}) with $d_0=t_3\cos(k)$, $d_x=t_0+t_1\cos(k)$, $d_y=t_2\cos(k)$ and $d_z=0$ at zero temperature.
Here, $t_0=1$, $t_1=6$, $t_2=2$ and $t_3=5$. Inset in (b): $t_3=-5$ so that the conductivity becomes negative.}
\label{fig2}
\end{figure}

We now focus on the longitudinal conductivity. It is a well-known fact that for a Hermitian system with symmetric energy spectra with
$\varepsilon(-{\bm k})=\varepsilon({\bm k})$, there does not exist a second-order nonlinear longitudinal conductivity. However, Eq.~(\ref{JEqn})
remarkably shows that
a second-order nonlinear longitudinal conductivity arises in a non-Hermitian system due to the geometric structures of wave functions, i.e.,
\begin{equation}
\chi_{\lambda\lambda\lambda}=-\frac{e^3\tau }{ 2(1+i\omega \tau) }\int_k f_0 \partial^2_\lambda\bar{A}_\lambda.
\end{equation}
In the dc limit, $\chi$ scales as $\tau$ instead of $\tau^2$. At high frequencies $\omega\tau \gg 1$ but below
the interband transition threshold, the prefactor
in $\chi$ is independent of the scattering time so that the nonlinear longitudinal conductivity directly measures
the geometric structure in the NHABC.

With time-reversal symmetry, the nonlinear longitudinal conductivity can be nonzero due to the fact that
$\bar{A}_{\lambda}(-{\bm k})=\bar{A}_{\lambda}({\bm k})$ enforced by time-reversal symmetry~\cite{Supplement}.
However, it is forced to vanish in an inversion symmetric system because inversion symmetry imposes constraints that
$\bar{A}_{\lambda}(-{\bm k})=-\bar{A}_{\lambda}({\bm k})$~\cite{Supplement}, making $f_0 \partial^2_\lambda\bar{A}_\lambda$ an odd function. In Fig.~\ref{fig2}(a), we plot $\chi=\chi_{xxx}$ for a 1D Hamiltonian with time-reversal symmetry, showing that a significant nonlinear
longitudinal conductivity arises when the Fermi energy is close to the band edge.

Besides the nonlinear longitudinal conductivity, we are surprised to find that the NHABC can induce
a linear longitudinal conductivity,
\begin{equation}
\sigma_{\lambda\lambda}^{\mathrm{NA}}=e^2\int_k f_0 \partial_\lambda\bar{A}_\lambda,
\end{equation}
which is independent of frequencies and the scattering time.
Due to the constraint imposed by time-reversal symmetry, this conductivity is forced to vanish in a time-reversal invariant system.
Instead, we consider a Hamiltonian that breaks time-reversal symmetry and exhibits antisymmetric $\bar{A}$; the
NHABC induced linear conductivity reaches maxima when
the Fermi surface is near the band edges as illustrated in Fig.~\ref{fig2}(b). There, the conductivity becomes negative when either $t_3<0$ or $q<1$.

Before closing this section, we wish to briefly discuss the NHABC induced Hall effects in two dimensions.
To have nonzero linear Hall effects, one has to break time-reversal symmetry. With time-reversal symmetry, the linear Hall effects are forced
to vanish, and the nonlinear Hall effects are attributed to $\bar{\Omega}_{\lambda\nu}$. To characterize the Hall current, we
define three Hall pseudovectors as
$d_\lambda^{(1)}=\epsilon_{\mu\nu}\chi_{\mu\nu\lambda}/2$ with $\epsilon_{\mu\nu}$ being the 2D Levi-Civita symbol,
$d_\lambda^{(2)}=\epsilon_{\mu\nu}\chi_{\mu\lambda\nu}/2$, and
$d_\lambda^{(3)}=\epsilon_{\mu\nu}\chi_{\lambda\mu\nu}/2$,
which contribute to the Hall current as ${\bm j}_{1}^0=({\bm{ \mathcal{E} }} \times {\bm e}_z) (\bm{ d}^{(1)}\cdot{\bm {\mathcal{E}} }^*)/2$,
${\bm j}_1^{2\omega}=({\bm{ \mathcal{E} }} \times {\bm e}_z) (\bm{ d}^{(1)}\cdot{\bm {\mathcal{E}} })/2$,
${\bm j}_2^0=({\bm{ \mathcal{E} }^*} \times {\bm e}_z) (\bm{ d}^{(2)}\cdot{\bm {\mathcal{E}} })/2$,
${\bm j}_2^{2\omega}=({\bm{ \mathcal{E} }} \times {\bm e}_z) (\bm{ d}^{(2)}\cdot{\bm {\mathcal{E}} })/2$,
${\bm j}_3^{0}={\bm d}^{(3)} [(\bm{\mathcal{E}\times \bm{\mathcal{E}}^* })\cdot {\bm e}_z]/2$ and
${\bm j}_3^{2\omega}=0$.
The NHABC cannot yield nonzero ${\bm d}^{(1)}$ since
$\partial_{\mu\nu}\bar{A}_\lambda=\partial_{\nu\mu}\bar{A}_\lambda$.

In summary, we have discovered anomalous coherent oscillations of a wave packet
induced by the NHABC.
While we demonstrate our prediction
in a pseudo-Hermitian Hamiltonian,
the anomalous oscillations in an ac electric field may also be observed in other
non-Hermitian systems~\cite{Supplement}, such as a system with skin effects~\cite{Yao2018PRL1,Xiong2018JPC},
given that the dynamics of a wave packet in a non-Hermitian Hamiltonian is independent
of boundary conditions~\cite{Pengfei2021Arixv}.
In the Supplementary Material, we also propose a practical scheme with coupled resonator optical waveguides~\cite{Hafezi2011NP,Fan2004PRL,Longhi2015SR}
to observe the anomalous
oscillations. We further provide a generic theory, showing the existence of a NHABC induced anomalous linear longitudinal conductivity independent of the scattering
time in a time-reversal symmetry broken system and a second-order anomalous nonlinear longitudinal conductivity
in a time-reversal invariant system.
Given that non-Hermitian physics can widely exist in disordered or strongly correlated systems (the
conductivity may be insensitive to boundary conditions even for a system with skin effects~\cite{Sato2021PRL}),
the anomalous longitudinal conductivities may be observed in these materials.
Our work thus opens a new direction for studying anomalous transport phenomena induced by NHABC in non-Hermitian systems.

\begin{acknowledgments}
We thank T. Qin and Y.-B. Yang for helpful discussions.
The work is
supported by the National Natural Science Foundation
of China (Grant No. 11974201) and the start-up fund from Tsinghua University.
\end{acknowledgments}

\begin{widetext}

\setcounter{equation}{0} \setcounter{figure}{0} \setcounter{table}{0}
\renewcommand{\theequation}{S\arabic{equation}} \renewcommand{\thefigure}{S\arabic{figure}} \renewcommand{\bibnumfmt}[1]{[S#1]}
\renewcommand{
\citenumfont}[1]{S#1}

In the supplementary material, we will show that the velocity of a wave packet is dependent on the Berry curvature defined
by right eigenstates in non-Hermitian systems in Section S-1,
present the eigenvalues of the Hamiltonian in the main text in Section S-2,
discuss the effects of non-Hermitian anomalous Berry connection
in a flattened Hamiltonian in Section S-3,
analyze the constraints on non-Hermitian anomalous Berry connection imposed by time-reversal symmetry
and inversion symmetry or two-fold rotational symmetry in Section S-4,
present non-Hermitian anomalous Berry connection in other non-Hermitian models with skin effects in Section S-5,
and
provide an experimental scheme with coupled resonator optical waveguides in Section S-6.

\section{S-1. Relevant Berry curvature in non-Hermitian systems}
In this section, we will show that the velocity of a wave packet is dependent on the Berry curvature defined
by right eigenstates.

\subsection{A. Expectation value of an observable}
A quantum non-Hermitian system usually appears due to the system coupled to an environment
constituting an open quantum system. Such a system is usually described by the master equation
\begin{equation}
i\partial_t \rho=[H_{eff},\rho]^\prime+\sum_jL_j\rho L_j^\dagger,
\end{equation}
where $\rho$ is the density matrix, $[H_{eff},\rho]^\prime=H_{eff}\rho-\rho H_{eff}^\dagger$ with $H_{eff}=H_s-i\frac{1}{2}\sum_jL_j^\dagger L_j$
being an effective non-Hermitian Hamiltonian,
$H_s$ is the Hamiltonian of the system, and $L_j$ is the jump operator. If we consider short time dynamics or consider the postselection so that the results corresponding to occurrence of a quantum jump are discarded, the dynamics is governed by
\begin{equation}
i\partial_t \rho=[H_{eff},\rho]^\prime.
\end{equation}
If an initial state is a pure state
$|\Psi(t=0)\rangle$ [$\rho(t=0)=|\Psi(t=0)\rangle \langle \Psi(t=0)|$], then the state evolves as
$\rho(t)=|\Psi(t)\rangle \langle \Psi(t)|$,
where $|\Psi(t)\rangle$ evolves according to
\begin{equation}
i\partial_t |\Psi(t)\rangle=H_{eff} |\Psi (t)\rangle.
\end{equation}
The expectation value of an observable $O$ is thus given by
\begin{equation}
\langle O \rangle=\frac{{\text{Tr}}[O\rho(t)]}{{\text{Tr}}[\rho(t)]}
=\frac{{\text{Tr}}[O|\Psi (t)\rangle \langle \Psi (t)|]}{{\text{Tr}}[|\Psi (t)\rangle \langle \Psi (t)|]}
=\langle \Psi_1|O|\Psi_1\rangle,
\end{equation}
where $|\Psi_1\rangle=|\Psi(t)\rangle/\sqrt{\langle \Psi(t) |\Psi(t)\rangle}$.
In the derivation, we have chosen an orthonormal basis including $|\Psi_1\rangle$ for the Hilbert space,
$\beta=\{|\Psi_1\rangle,|\Psi_2\rangle,\cdots,|\Psi_N\rangle\}$ ($N$ denotes the dimension of the Hilbert space),
so that
\begin{equation}
{\text{Tr}}[O|\Psi (t)\rangle \langle \Psi (t)|]=\langle \Psi_1|O| \Psi_1\rangle \langle \Psi(t) |\Psi(t)\rangle +\sum_{j=2}^N
\langle \Psi_j | O|\Psi (t)\rangle \langle \Psi (t)| \Psi_j\rangle =\langle \Psi_1|O| \Psi_1\rangle \langle \Psi(t) |\Psi(t)\rangle.
\end{equation}
This tells us that the expectation value is completely determined by the evolving state $|\Psi(t)\rangle$.

In a strongly correlated or disordered system, it has been justified in Ref.~\cite{Michishita2020PRLS} that the effective non-Hermitian Hamiltonian in the single-particle Green's function corresponds to the non-Hermitian Hamiltonian in the context of an open quantum system under
postselection, indicating that the expectation value of an observable of a wave packet is still determined by the evolving
state $|\Psi(t)\rangle$ under the effective non-Hermitian Hamiltonian.

\subsection{B. Derivation of the semiclassical equations of motion}
We now provide the detailed derivation of the semiclassical dynamics of Bloch electrons (a brief version can be found in the
Supplementary Material in Ref.~\cite{Xu2017PRLS}). Consider a wave packet as an initial state, which can be written
in terms of the right Bloch eigenstates of
a non-Hermitian Hamiltonian $H$ without an external force as
\begin{equation}
|\Phi(t=0)\rangle=\int_{k}a(\bm{k},t=0)|\phi_{n}^{R}(\bm{k})\rangle,
\end{equation}
where $|\phi_{n}^{R}(\bm{k})\rangle=e^{i{\bm k}\cdot{\bm r}}|u_{n}^{R}(\bm{k})\rangle$
is the Bloch state with $|u_{n}^{R}(\bm{k})\rangle$ being the right
eigenstate of a non-Hermitian Hamiltonian.
The wave packet evolves with time as
\begin{equation}
|\Phi(t)\rangle=\int_{k}a(\bm{k},t)|\phi_{n}^{R}(\bm{k})\rangle,\label{eq:WavePacket-1}
\end{equation}
where $a({\bm k},t)=|a({\bm k},t)|e^{i\gamma({\bm k},t)}$
with the amplitude $|a({\bm k},t)|$ taking the Gaussian form centered
at $\bm{k}_{c}=\int_{k}|a(\bm{k},t)|^{2}\bm{k}/\int_{k}|a(\bm{k},t)|^{2}=\int_{k}|a^{\prime}(\bm{k},t)|^{2}\bm{k}$
where $|a^{\prime}(\bm{k},t)|^{2}\equiv|a(\bm{k},t)|^{2}/\int_{k^{\prime}}|a(\bm{k}^{\prime},t)|^{2}$
(while $|a({\bm k},t)|$ may decrease with time in the presence of
a loss term so that $\langle\Phi|\Phi\rangle=\int_{k}|a(\bm{k},t)|^{2}<1$,
we can evaluate the expectation value of an observable by mutiplying
a factor $1/\langle\Phi|\Phi\rangle$).

Let us first derive the expression of $\gamma({\bm k},t)$ which
contains both dynamical and geometric phases. In the presence of an
external force, the Schr{\"o}dinger equation reads
\begin{equation}
i\partial_{t}|\psi(t)\rangle=(H-\bm{F}\cdot\hat{\bm{r}})|\psi(t)\rangle.
\end{equation}
Consider a Bloch eigenstate at $\bm{k}_{0}$ as an initial state at time $t_0$ and
suppose that the external force is sufficiently weak so that the evolving
state is still a Bloch eigenstate multiplied by a factor, that is,
\begin{align}
|\psi(t)\rangle & =\alpha(t)|\phi_{n\bm{k}(t)}^{R}\rangle=\alpha(t)e^{i\bm{k}(t)\cdot\bm{r}}|u_{n\bm{k}(t)}^{R}\rangle\\
\bm{k}(t) & =\bm{k}_{0}+\bm{F}t.
\end{align}
We now substitute $|\psi(t)\rangle$ into the Schr{\"o}dinger equation,
yielding
\begin{align}
i\partial_{t}|\psi(t)\rangle= & ie^{i\bm{k}(t)\cdot\bm{r}}|u_{n}^{R}(\bm{k}(t))\rangle\partial_{t}\alpha(t)-\alpha(t)e^{i\bm{k}(t)\cdot\bm{r}}|u_{n}^{R}(\bm{k}(t))\rangle\bm{F}\cdot\bm{r}+i\alpha(t)e^{i\bm{k}(t)\cdot\bm{r}}\partial_{t}|u_{n}^{R}(\bm{k}(t))\rangle\\
= & (H-\bm{F}\cdot\hat{\bm{r}})|\psi(t)\rangle\\
= & \alpha(t)\hat{H}e^{i\bm{k}(t)\cdot\bm{r}}|u_{n}^{R}(\bm{k}(t))\rangle-\alpha(t)e^{i\bm{k}(t)\cdot\bm{r}}|u_{n}^{R}(\bm{k}(t))\rangle\bm{F}\cdot\bm{r}\\
= & \alpha(t)\varepsilon_{n}(\bm{k}(t))e^{i\bm{k}(t)\cdot\bm{r}}|u_{n}^{R}(\bm{k}(t))\rangle-\alpha(t)e^{i\bm{k}(t)\cdot\bm{r}}|u_{n}^{R}(\bm{k}(t))\rangle\bm{F}\cdot\bm{r}.
\end{align}
We thus obtain
\begin{equation} \label{EqAlpha}
i|u_{n}^{R}(\bm{k}(t))\rangle\partial_{t}\alpha(t)=
\alpha(t)\varepsilon_{n}(\bm{k}(t))|u_{n}^{R}(\bm{k}(t))\rangle-i\alpha(t)\partial_{t}|u_{n}^{R}(\bm{k}(t))\rangle.
\end{equation}
To derive the equation satisfied by $\alpha(t)$, we consider any state written as a linear combination of
the left eigenstates $|u_{j}^{L}(\bm{k}(t))\rangle$ [note that $\{|u_{j}^{L}(\bm{k}(t))\rangle\}$ constitute an ordered basis, although it may not be orthogonal], that is,
\begin{equation}
|u^\prime_{\bm{k}(t)}\rangle =c_n^* |u_{n}^{L}(\bm{k}(t))\rangle+\sum_{j \neq n} c_j^* |u_{j}^{L}(\bm{k}(t))\rangle.
\end{equation}
We then multiply Eq.~(\ref{EqAlpha}) by $\langle u^\prime_{\bm{k}(t)}|$, resulting
in
\begin{align}
ic_n\partial_{t}\alpha(t) & =c_n\alpha(t)[\varepsilon_{n}(\bm{k}(t))-i \langle u_{n}^{L}(\bm{k}(t))|\partial_{t}|u_{n}^{R}(\bm{k}(t))\rangle ]
-i\alpha(t)\sum_{j\neq n} c_j \langle u_{j}^{L}(\bm{k}(t))|\partial_{t}|u_{n}^{R}(\bm{k}(t))\rangle,
\end{align}
where we have used the biorthogonal relation $\langle u_m^{L}({\bm k})|u_n^{R}({\bm k})\rangle =\delta_{mn}$.
The final term vanishes because multiplying Eq.~(\ref{EqAlpha}) by $\sum_{j \neq n} c_j \langle u_{j}^{L}(\bm{k}(t))|$ yields
\begin{equation}
\alpha(t)\sum_{j\neq n} c_j \langle u_{j}^{L}(\bm{k}(t))|\partial_{t}|u_{n}^{R}(\bm{k}(t))\rangle=0.
\end{equation}
Thus, we obtain
\begin{align} \label{EqAlpha2}
i\partial_{t}\alpha(t) & =\alpha(t)\varepsilon_{n}(\bm{k}(t))-i\alpha(t)\langle u_{n}^{L}(\bm{k}(t))|\partial_{t}|u_{n}^{R}(\bm{k}(t))\rangle=\alpha(t)\left[\varepsilon_{n}(\bm{k}(t))-\tilde{{\bm A}}_{n{\bm k}(t)}\cdot{\bm F}\right],
\end{align}
yielding
\begin{equation} \label{BerryPhase}
\alpha(t)=e^{-i\int_{t_{0}}^{t}dt^{\prime}\left[\varepsilon_{n}(\bm{k}(t^{\prime}))-\tilde{{\bm A}}_{n}(\bm{k}(t^{\prime}))\cdot{\bm F}\right]},
\end{equation}
where $\tilde{{\bm A}}_{n}(\bm{k}(t))=i\langle u_{n}^{L}(\bm{k})|\partial_{\bm{k}}u_{n}^{R}(\bm{k})\rangle|_{\bm{k}=\bm{k}(t)}$
is the left-right Berry connection contributing to the geometric phase.

We note that in the derivation, while one can also use the right eigenvectors
to generate $|u^\prime_{\bm{k}(t)}\rangle$, that is,
\begin{equation}
|u^\prime_{\bm{k}(t)}\rangle =b_n^* |u_{n}^{R}(\bm{k}(t))\rangle+\sum_{j \neq n} b_j^* |u_{j}^{R}(\bm{k}(t))\rangle,
\end{equation}
multiplying Eq.~(\ref{EqAlpha}) by this expansion cannot give us an equation as concise as Eq.~(\ref{EqAlpha2})
due to the absence of an orthogonal relation for right eigenvectors in non-Hermitian systems
[$\langle u_m^{R}({\bm k})|u_n^{R}({\bm k})\rangle$ usually does not vanish if $m\neq n$].
Note here we consider a system without exceptional points; otherwise, neither right eigenvectors nor left eigenvectors
can constitute a basis for the Hilbert space.

We are now in a position to derive the center of mass of the wave packet in real space as time evolves based on
Eq. (\ref{eq:WavePacket-1}); the location of the wave packet is given by
\begin{align}
&{\bm r}_{c}(t) \nonumber \\
  =&\langle\Phi(t)|\hat{{\bm r}}|\Phi(t)\rangle/\langle\Phi(t)|\Phi(t)\rangle \\
  =&\int d{\bm r}\int_{k}\int_{k^{\prime}}|a({\bm k}^{\prime},t)||a({\bm k},t)|e^{i\gamma({\bm k},t)}e^{-i\gamma({\bm k}^{\prime},t)}e^{-i{\bm k}^{\prime}\cdot{\bm r}} [u_{n{\bm k}^{\prime}}^{R}({\bm r})]^\dagger {\bm r}e^{i{\bm k}\cdot{\bm r}}u_{n{\bm k}}^{R}(\bm{r})
  /\langle\Phi|\Phi\rangle \\
  =&\int d{\bm r}\int_{k}\int_{k^{\prime}}|a({\bm k}^{\prime},t)||a({\bm k},t)|e^{i\gamma({\bm k},t)}e^{-i\gamma({\bm k}^{\prime},t)}e^{i({\bm k}-{\bm k}^{\prime})\cdot{\bm r}}{\bm r}[u_{n{\bm k}^{\prime}}^{R}(\bm{r})]^{\dagger}u_{n{\bm k}}^{R}(\bm{r})/\langle\Phi|\Phi\rangle \\
  =&\int_{k}\int_{k^{\prime}}\int d{\bm r}|a({\bm k}^{\prime},t)||a({\bm k},t)|e^{i\gamma({\bm k},t)}e^{-i\gamma({\bm k}^{\prime},t)}[u_{n{\bm k}^{\prime}}^{R}(\bm{r})]^{\dagger}u_{n{\bm k}}^{R}(\bm{r})(-i\partial_{{\bm k}})e^{i({\bm k}-{\bm k}^{\prime})\cdot{\bm r}}/\langle\Phi|\Phi\rangle \\
  =&\int_{k}\int_{k^{\prime}}\int d{\bm r}e^{i({\bm k}-{\bm k}^{\prime})\cdot{\bm r}}|a({\bm k}^{\prime},t)|e^{-i\gamma({\bm k}^{\prime},t)}[u_{n{\bm k}^{\prime}}^{R}(\bm{r})]^{\dagger}i\partial_{{\bm k}}|a({\bm k},t)|e^{i\gamma({\bm k},t)}u_{n{\bm k}}^{R}(\bm{r})/\langle\Phi|\Phi\rangle \\
  =&\int_{k}\int d{\bm r}|a({\bm k},t)|e^{-i\gamma({\bm k},t)}[u_{n{\bm k}}^{R}(\bm{r})]^{\dagger}i\partial_{{\bm k}}|a({\bm k},t)|e^{i\gamma({\bm k},t)}u_{n{\bm k}}^{R}(\bm{r})/\langle\Phi|\Phi\rangle \\
  =&\int_{k}\int d{\bm r}|a({\bm k},t)|e^{-i\gamma({\bm k},t)}[u_{n{\bm k}}^{R}(\bm{r})]^{\dagger} \left[e^{i\gamma({\bm k},t)}u_{n{\bm k}}^{R}(\bm{r})i\partial_{{\bm k}}|a({\bm k},t)|-(\partial_{{\bm k}}\gamma({\bm k},t))|a({\bm k},t)|e^{i\gamma({\bm k},t)}u_{n{\bm k}}^{R}(\bm{r})+\right.\\
 &\left.
 |a({\bm k},t)|e^{i\gamma({\bm k},t)}(i\partial_{{\bm k}})u_{n{\bm k}}^{R}(\bm{r})\right]/\langle\Phi|\Phi\rangle \\
  =&\int_{k}|a({\bm k},t)|i\partial_{{\bm k}}|a({\bm k},t)|/\langle\Phi|\Phi\rangle-\int_{k}|a({\bm k},t)|^{2}(\partial_{{\bm k}}\gamma({\bm k},t))/\langle\Phi|\Phi\rangle+\int_{k}|a({\bm k},t)|^{2}\int d{\bm r}[u_{n{\bm k}}^{R}(\bm{r})]^{\dagger}i\partial_{{\bm k}}u_{n{\bm k}}^{R}(\bm{r})/\langle\Phi|\Phi\rangle \\
  =&-\int_{k}|a^{\prime}({\bm k},t)|^{2}\partial_{{\bm k}}\gamma({\bm k},t)+\int_{k}|a^{\prime}({\bm k},t)|^{2}\int d{\bm r}[u_{n{\bm k}}^{R}(\bm{r})]^{\dagger}i\partial_{{\bm k}}u_{n{\bm k}}^{R}(\bm{r})\\
  =&-\int_{k}|a^{\prime}({\bm k},t)|^{2}\partial_{{\bm k}}\gamma({\bm k},t)+\int_{k}|a^{\prime}({\bm k},t)|^{2}\langle u_{n\bm{k}}^{R}|i\partial_{\bm{k}}|u_{n\bm{k}}^{R}\rangle\\
  \approx &-\frac{\partial\gamma({\bm k}_{c},t)}{\partial{\bm k}_{c}}+i\langle u_{n{\bm k}_{c}}^{R}|\partial_{{\bm k}_{c}}|u_{n{\bm k}_{c}}^{R}\rangle \\
  =&-\frac{\partial\gamma({\bm k}_{c},t)}{\partial{\bm k}_{c}}+{\bm A}_{n}({\bm k}_{c}),
\end{align}
where ${\bm A}_{n}({\bm k}_{c})=i\langle u_{n{\bm k}_{c}}^{R}|\partial_{{\bm k}_{c}}|u_{n{\bm k}_{c}}^{R}\rangle$
is the right-right Berry connection with $|u_{n{\bm k}_{c}}^{R}\rangle\equiv|u_{n}^{R}({\bm k}_{c})\rangle$.
With the aid of Eq.~(\ref{BerryPhase}), we have
\[
\gamma({\bm k}_{c},t)=-\int_{t_{0}}^{t}dt^{\prime}\text{Re}\left[\varepsilon_{n}({\bm k}(t^{\prime}))-\tilde{{\bm A}}_{n}({\bm k}(t^{\prime}))\cdot{\bm F}\right].
\]
We thus can derive the velocity of the wave packet as
\begin{align*}
{\bm v}_{c}(t_{0}) & =\frac{{\bm r}_{c}(t_{0}+\delta t)-{\bm r}_{c}(t_{0})}{\delta t}\\
 & =\partial_{{\bm k}_{c}}\frac{\delta t\left[\text{Re}[\varepsilon_{n}({\bm k}_{c})]-\text{Re}[\tilde{{\bm A}}_{n}({\bm k}_{c})]\cdot{\bm F}\right]}{\delta t}+\frac{d{\bm A}_{n}({\bm k}_{c})}{dt}|_{t=t_{0}}\\
 & =\partial_{{\bm k}_{c}}\text{Re}[\varepsilon_{n}({\bm k}_{c})]-\partial_{{\bm k}_{c}}\text{Re}[\tilde{A}_{n,\nu}({\bm k}_{c})]\cdot F_{\nu}+\frac{d{\bm A}_{n}({\bm k}_{c})}{dt}|_{t=t_{0}}\\
 & =\partial_{{\bm k}_{c}}\text{Re}[\varepsilon_{n}({\bm k}_{c})]-(\partial_{{\bm k}_{c}}\text{Re}[\tilde{A}_{n,\nu}({\bm k}_{c})])\cdot\overset{\cdot}{k}_{c,\nu}+\frac{d{\bm A}_{n}({\bm k}_{c})}{dt}|_{t=t_{0}},\\
 & =\partial_{{\bm k}_{c}}\text{Re}[\varepsilon_{n}({\bm k}_{c})]-\partial_{{\bm k}_{c}}A_{n,\nu}({\bm k}_{c})\cdot\overset{\cdot}{k}_{c,\nu}+\frac{d{\bm A}_{n}({\bm k}_{c})}{dt}|_{t=t_{0}}+\partial_{{\bm k}_{c}}\left[A_{n,\nu}({\bm k}_{c})-\text{Re}\tilde{A}_{n,\nu}({\bm k}_{c})\right]\cdot\overset{\cdot}{k}_{c,\nu}\\
 & =\partial_{{\bm k}_{c}}\text{Re}[\varepsilon_{n}({\bm k}_{c})]-\overset{\cdot}{{\bm k}}_{c}\times\bm{\Omega}+\partial_{{\bm k}_{c}}\text{Re}\left[A_{n,\nu}({\bm k}_{c})-\tilde{A}_{n,\nu}({\bm k}_{c})\right]\cdot\overset{\cdot}{k}_{c,\nu}\\
 & =\partial_{{\bm k}_{c}}\text{Re}[\varepsilon_{n}({\bm k}_{c})]-\overset{\cdot}{{\bm k}}_{c}\times\bm{\Omega}+\partial_{{\bm k}_{c}}\bar{A}_{n,\nu}({\bm k}_{c})\cdot\overset{\cdot}{k}_{c,\nu}\\
 & =\partial_{{\bm k}_{c}}[\text{Re}[\varepsilon_{n}({\bm k}_{c})]+\bar{A}_{n,\nu}({\bm k}_{c})\cdot\overset{\cdot}{k}_{c,\nu}]-\overset{\cdot}{{\bm k}}_{c}\times\bm{\Omega},
\end{align*}
where $\bar{A}_{n,\nu}({\bm k}_{c})\equiv\text{Re}\left[A_{n,\nu}({\bm k}_{c})-\tilde{A}_{n,\nu}({\bm k}_{c})\right]$
and $\bm{\Omega}=i\langle\nabla_{{\bm k}}u_{n}^{R}({\bm k})|\times|\nabla_{{\bm k}}u_{n}^{R}({\bm k})\rangle$.
In the derivation, we have used the fact that ${\bm F}=\overset{\cdot}{{\bm k}}$.
We can clearly see that the velocity of a wave packet is dependent on the right-right
Berry curvature, which is independent of the left eigenstates.

\section{S-2. Eigenenergies of the Hamiltonian}
The Hamiltonian (5) in the main text is pseudo-Hermitian, i.e., $U^{-1}HU=H^\dagger$ with $U=\text{diag}(\sqrt{q},1/\sqrt{q})$, and its eigenenergies are real,
$
\varepsilon_{\pm}=mc_0+\xi_\pm
$
with $\xi_{\pm}=\pm \sqrt{ab(d_x^2+d_y^2)+m^2c_1^2}$ and $c_0=(1-q^2)/(2q)$ and $c_1=(1+q^2)/(2q)$.

\section{S-3. Effects of non-Hermitian anomalous Berry connection in a flattened Hamiltonian}
To illustrate the non-Hermitian anomalous Berry connection effects, we can take a limit and consider a flattened Hamiltonian
\begin{equation}
H_{\mathrm{F}}=|{u_+^R}\rangle \langle {u_+^L}|-|{u_-^R}\rangle \langle {u_-^L}|,
\end{equation}
where $|{u_{\pm}^R}\rangle$ and $\langle{u_{\pm}^L}|$ are the normalized right and left eigenstates of the
Hamiltonian~(5), respectively, corresponding to eigenenergy $\varepsilon_{\pm}$.
For the flattened Hamiltonian, its eigenenergies are $\varepsilon_{\mathrm{F},\pm}=\pm 1$,
which is independent of quasimomentum $k$ and thus the group velocity contributed by the energy
dispersion vanishes. However, the non-Hermitian anomalous velocity remains the same as in the
Hamiltonian~(5), since it only depends on the wave functions. In this case, we can clearly
see that even though the traditional semiclassical equations of motion predict the absence of motion for an electron
in the presence of an external electric field, our theory indicates that the electron can still move due to
the emergence of a velocity arising from the non-Hermiticity and local geometric effects.
More interestingly, given that ${\bm v}_{\mathrm{NA}}$ is proportional to the electric field, one can
control the direction of the velocity of an electron by suddenly reversing the direction of the
electric field.

\section{S-4. Symmetry constraints on non-Hermitian anomalous Berry connection}
In this section, we explore the constraints on non-Hermitian anomalous Berry connection imposed by time-reversal
symmetry and inversion or two-fold rotational symmetry.

\subsection{A. Time-reversal symmetry}
For a non-Hermitian system, time-reversal and particle-hole symmetries develop into four categories~\cite{Sato2019PRXS,HZhou2019PRBS}:
\begin{align}
T_{\pm}H^{*}(-\bm{k})T_{\pm}^{\dagger} & =\pm H(\bm{k})\\
\Xi_{\pm}H^{T}(-\bm{k})\Xi_{\pm}^{\dagger} & =\pm H(\bm{k}),
\end{align}
where $T_{\pm}$ and $\Xi_{\pm}$ are unitary operators. The Hamiltonian (5) in the main text respects the
time-reversal symmetry with $T_+$ being an identity matrix. We thus only consider the constraint imposed by
$T_+$ with $T_+T_+^*=1$.

Let $|u_{n}^{R}({\bm k})\rangle$ and $\langle u_n^{L}({\bm k})|$ be the normalized right and left eigenstate of
a non-Hermitian Hamiltonian $H({\bm k})$ in the $n$th band corresponding to an eigenvalue
$\varepsilon_n({\bm k})$, respectively, that is,
$H({\bm k})|u_n^{R}({\bm k})\rangle=\varepsilon_{k}|u_n^{R}({\bm k})\rangle$ and
$\langle u_n^{L}({\bm k})|H({\bm k})=\varepsilon_{k}\langle u_n^{L}({k})|$.
Based on the symmetry constraint,
one can easily obtain
\begin{align}
H(-{\bm k})T_{+}|u_n^{R}({\bm k})\rangle^{*} & =T_{+}T_{+}^{\dagger}H(-{\bm k})T_{+} |u_n^{R}({\bm k})\rangle^{*}
=T_{+}H^{*}({\bm k}) |u_n^{R}({\bm k})\rangle^{*} =\varepsilon_{n}^{*}({\bm k})T_{+} |u_n^{R}({\bm k})\rangle^{*},
\end{align}
and
\begin{align}
(\langle u_n^{L}({\bm k})|)^{*}T_{+}^{\dagger}H(-{\bm k}) =(\langle u_n^{L}({\bm k})|)^{*} H^{*}({\bm k})T_{+}^{\dagger}
=\varepsilon_{n}^{*}({\bm k})(\langle u_n^{L}({\bm k})|)^{*} T_{+}^{\dagger},
\end{align}
indicating that $T_{+}|u_n^{R}({\bm k})\rangle^{*}$ and $(\langle u_n^{L}({\bm k})|)^{*}T_{+}^{\dagger}$ are the right
and left eigenstates of $H(-{\bm k})$ with eigenvalue $\varepsilon_{n}^{*}({\bm k})$, respectively.
Since non-Hermitian anomalous Berry connection is gauge independent, we choose a gauge so that
$|u_n^{R}(-{\bm k})\rangle=T_{+}|u_n^{R}({\bm k})\rangle^{*}$
and $\langle u_n^{L}(-{\bm k})|=(\langle u_n^{L}({\bm k})|)^{*}T_{+}^{\dagger}$,
which fulfill the normalization condition $\langle u_n^{R}(-{\bm k}) |u_n^{R}(-{\bm k})\rangle=1$ and
$\langle u_n^{L}(-{\bm k}) |u_n^{R}(-{\bm k})\rangle=1$.
We now derive the constraints as
\begin{align}
A_{n,\mu}(-{\bm k}) & =-i\langle u_n^{R}(-{\bm k})|\partial_{\mu}u_n^{R}(-{\bm k})\rangle\\
 & =-i(T_{+}|u_n^{R}({\bm k})\rangle^{*})^{\dagger}T_{+}|\partial_{\mu}u_n^{R}({\bm k})\rangle^{*}\\
 & =-i(\langle u_n^{R}({\bm k})|)^{*} |\partial_{\mu}u_n^{R}({\bm k})\rangle^{*}\\
 & =(A_{n,\mu}({\bm k}))^{*},
\end{align}
and
\begin{align}
\tilde{A}_{n,\mu}(-{\bm k}) & =-i\langle u_n^{L}(-{\bm k})|\partial_{\mu}u_n^{R}(-{\bm k})\rangle
 =-i\langle u_n^{L}({\bm k})|^{*}T_{+}^{\dagger}T_{+}\partial_{\mu}|u_n^{R}({\bm k})\rangle^{*}
 =(\tilde{A}_{n,\mu}({\bm k}))^{*},
\end{align}
which yield
\begin{equation}
\bar{\bm A}_{n}(-{\bm k})=\bar{\bm A}_{n}({\bm k}).
\end{equation}
We thus conclude that the time-reversal symmetry $T_{+}$ with $T_{+}T_{+}^{*}=1$ ensures that
non-Hermitian Berry connection $\bar{\bm A}$ is an even function with respect to ${\bm k}$.

\subsection{B. Inversion symmetry or two-fold rotational symmetry}
We now consider inversion symmetry or two-fold rotational symmetry which forces the Hamiltonian to respect
\begin{equation}
UH({-{\bm k}})U^{\dagger}=H({\bm k}),
\end{equation}
where $U$ is a unitary operator. One can easily find that $U^{\dagger}|u_n^{R}({\bm k})\rangle$
and $\langle u_n^{L}({\bm k})|U$ are the right and left eigenstates of $H(-{\bm k})$
corresponding to eigenenergy $\varepsilon_{n}({\bm k})$, respectively. By choosing a gauge so that
$|u_n^R(-{\bm k})\rangle=U^{\dagger}|u_n^{R}({\bm k})\rangle$ and
$\langle u_n^L(-{\bm k}) |=\langle u_n^{L}({\bm k})|U$, one can derive that
\begin{eqnarray}
{\bm A}_n(-{\bm k})&=&-{\bm A}_n({\bm k}) \\
\tilde{{\bm A}}_n(-{\bm k})&=&-\tilde{{\bm A}}_n({\bm k}),
\end{eqnarray}
leading to
\begin{equation}
\bar{{\bm A}}_n(-{\bm k})=-\bar{{\bm A}}_n({\bm k}).
\end{equation}
This tells us that inversion symmetry maintains that non-Hermitian anomalous Berry connection is
antisymmetric with respect to ${\bm k}$. If a system has both time-reversal symmetry $T_+$ and
inversion or two-fold rational symmetry, then non-Hermitian anomalous Berry connection is
forced to vanish.

\subsection{C. $C_2\mathcal{T}$ or ${\mathcal{PT}}$ symmetry}
For the $C_2\mathcal{T}$ (product of two-fold rotational symmetry and time-reversal symmetry) or
${\mathcal{PT}}$ (product of inversion symmetry and time-reversal symmetry) symmetry, the Hamiltonian should satisfy
\begin{equation}
UT_{+}H({\bm k})^{*}(UT_{+})^{\dagger}=H({\bm k}),
\end{equation}
where $UT_{+}$ is a unitary operator. We consider the case that $(UT_{+})(UT_{+})^{*}=1$.
Similarly, one can find that $UT_+|u_n^{R}({\bm k})\rangle^*$
and $(\langle u_n^{L}({\bm k})|)^*(UT_+)^\dagger$ are the right and left eigenstates of $H({\bm k})$
corresponding to eigenenergy $\varepsilon_{n}^*({\bm k})$, respectively.
If a system is in a $C_2\mathcal{T}$ or ${\mathcal{PT}}$ symmetric state, that is,
$UT_{+}|u_n^{R}({\bm k})\rangle^{*}={\alpha}^*|u_n^{R}({\bm k})\rangle$
and $\langle u_n^{L}({\bm k})|^{*}(UT_{+})^{\dagger}=\alpha\langle u_n^{L}({\bm k})|$
with $|\alpha|=1$, then one can easily derive
\begin{eqnarray}
{A}_{n,\mu}({\bm k}) & =& i\alpha^{*}\partial_{\mu}\alpha-(A_{n,\mu}({\bm k}))^{*} \\
\tilde{A}_{n,\mu}({\bm k}) & =& i\alpha^{*}\partial_{\mu}\alpha-(\tilde{A}_{n,\mu}({\bm k}))^{*},
\end{eqnarray}
so that
\begin{equation}
\bar{{\bm A}}_{n}({\bm k})=-\bar{{\bm A}}_{n}({\bm k}),
\end{equation}
forcing non-Hermitian anomalous Berry connection to vanish.

\begin{figure}[t]
	\includegraphics[width = 0.8\linewidth]{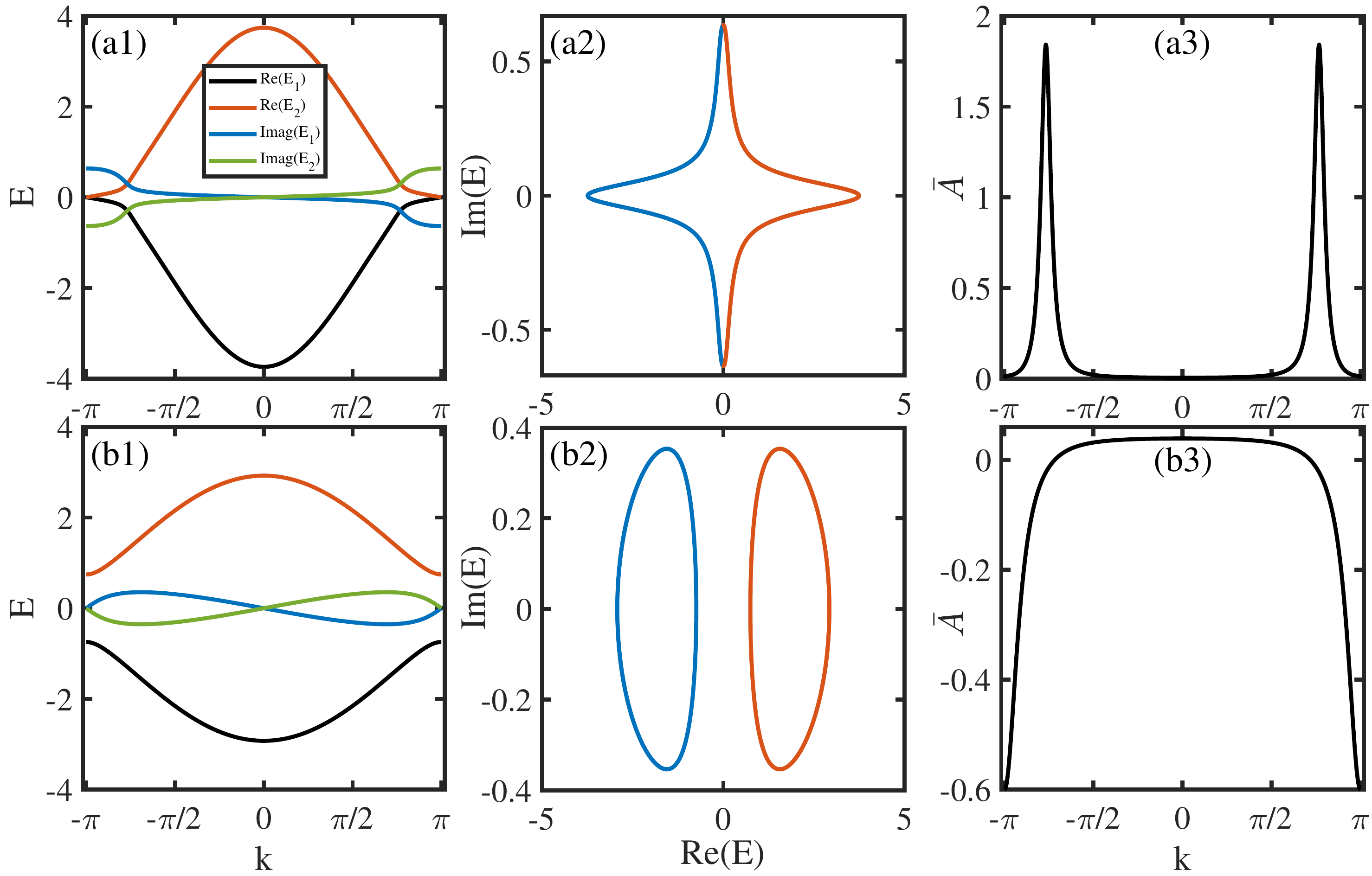}
	\caption{(Color online) (a1)(b1) Real and imaginary parts of energy spectra with respect to quasimomenta $k$ for the Hamiltonian (\ref{SkinHam})
with the corresponding energy spectra in the complex plane plotted in (a2) and (b2), respectively.
(a3)(b3) NHABC as a function of quasimomenta $k$. In (a1-a3), $t_1=2$, $t_2=1$, $t_3=0.8$ and $\gamma=4/3$,
and in (b1-b3), $t_1=2$, $t_2=1$, $t_3=0$ and $\gamma=4/3$.}	
	\label{figS1}
\end{figure}

\section{S-5. Non-Hermitian anomalous Berry connections in other models}
In the main text, we consider a model with pseudo-Hermiticity symmetry that exhibits large NHABC.
In fact, the NHABC is generically nonzero in a non-Hermitian system. Here, we show that NHABC can be
large in other well-known models with skin effects~\cite{Tony2016S,Yao2018PRL1S},
\begin{equation}\label{SkinHam}
H(k)=d_x\sigma_x+(d_y+i\gamma/2)\sigma_y,
\end{equation}
where $d_x=t_1+(t_2+t_3)\cos(k)$ and $d_y=(t_2-t_3)\sin(k)$ with $t_1$, $t_2$, $t_3$ and $\gamma$
being real parameters describing the corresponding hopping strength.

To show that NHABC has significant effects, we consider two typical cases with $t_1=2$, $t_2=1$, $t_3=0.8$ and $\gamma=4/3$,
and $t_1=2$, $t_2=1$, $t_3=0$ and $\gamma=4/3$. The system exhibits non-Hermitian skin effects under open
boundary conditions since the energy spectra in the complex plane in momentum space [see Fig.~\ref{figS1}(a2) and (b2)] have nonzero
winding. Figure~\ref{figS1}(a3) and (b3) further illustrate that the NHABC is also in the order of one in these systems, similar to
the case shown in Fig. 1(a) in the main text for a pseudo-Hermitian Hamiltonian.

\begin{figure}[t]
	\includegraphics[width = 0.8\linewidth]{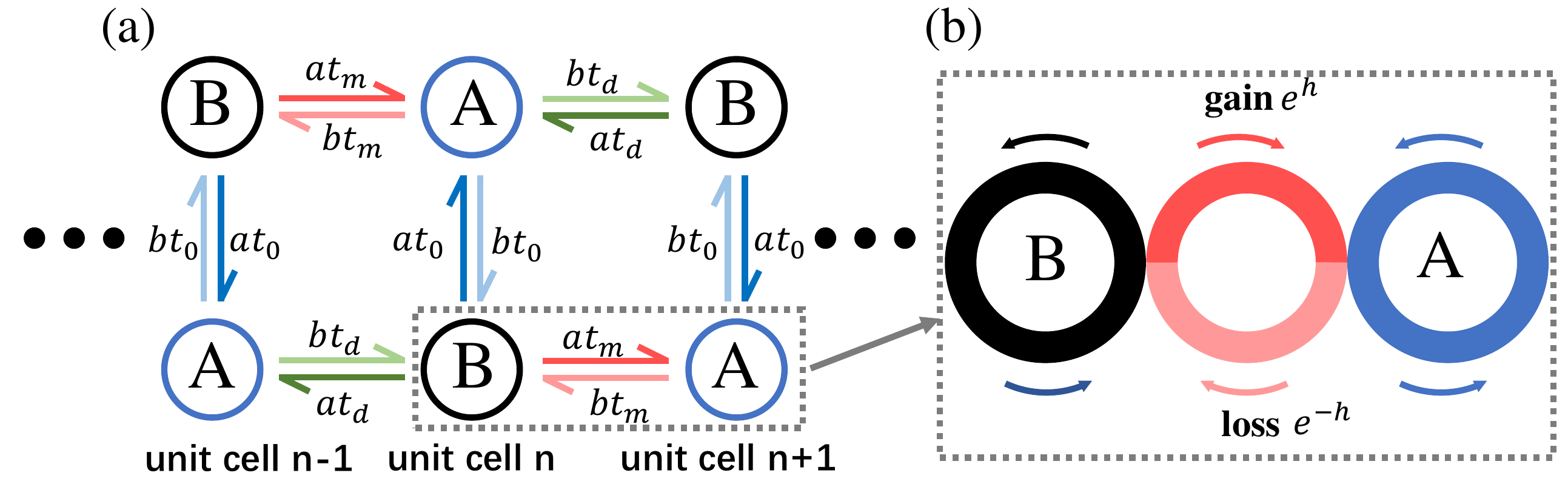}
	\caption{(Color online) (a) Lattice configurations to realize the Hamiltonian (5) in the main text.
Circles labeled by A and B represent site resonators. The effective hopping between the site resonators is
generated by a link waveguide (red circle) connecting two neighboring site resonators (black and blue circles) as shown in (b).
When gain and loss are involved in link waveguides, the required asymmetric hopping can be realized.}	
	\label{figS2}
\end{figure}

\section{S-6. An experimental scheme with coupled resonator optical waveguides}
To experimentally realize the Hamiltonian (5) with $d_z=d_0=0$, $d_x=t_0+t_1\cos k$ and $d_y=t_2\sin k$,
we first write down the Hamiltonian in real space
\begin{eqnarray}
H=\sum_n & \left[ (\begin{array}{cc}
                  |A_n\rangle & |B_n\rangle
                \end{array})
                \left(
                   \begin{array}{cc}
                     0 & at_0 \\
                     bt_0 & 0 \\
                   \end{array}
                 \right)
                \left(\begin{array}{c}
                  \langle A_n| \\
                  \langle B_n|
                \end{array}\right)
+
(\begin{array}{cc}
                  |A_n\rangle & |B_n\rangle
                \end{array})
                \left(
                   \begin{array}{cc}
                     0 & at_d \\
                     bt_m & 0 \\
                   \end{array}
                 \right)
                \left(\begin{array}{c}
                  \langle A_{n+1}| \\
                  \langle B_{n+1}|
                \end{array}\right)  \right.
\nonumber \\
& \left. +
(\begin{array}{cc}
                  |A_{n+1}\rangle & |B_{n+1}\rangle
                \end{array})
                \left(
                   \begin{array}{cc}
                     0 & at_m \\
                     bt_d & 0 \\
                   \end{array}
                 \right)
                \left(\begin{array}{c}
                  \langle A_{n}| \\
                  \langle B_{n}|
                \end{array}\right)
\right],
\end{eqnarray}
where $|A_n\rangle$ and $|B_n\rangle$ denote the state at site A and B in the $n$th unit cell, respectively, $t_m=(t_1+t_2)/2$ and $t_d=(t_1-t_2)/2$. To realize the Hamiltonian in coupled-resonator optical waveguides (CROWs), we consider a configuration for
site resonators and link waveguides as illustrated in Fig.~\ref{figS2}(a), where
A and B sites are interchanged in positions between two neighboring unit cells to realize the next-nearest neighbor
hopping between two neighboring unit cells. The hopping between two neighboring sites are implemented by coupling two site
resonators (denoted by black and blue circles) with a link waveguide (denoted by a red circle), as shown in Fig.~\ref{figS2}(b).
In the non-Hermitian case, the asymmetric hopping between two sites can be realized by
applying either gain $e^h$ or loss $e^{-h}$ over half circle for a light travelling in a link waveguide~\cite{Longhi2015SRS}.
For instance, consider a light propagating counterclockwise in a site resonator and
a gain in the upper semicircle and a loss in the lower semicircle in a link waveguide (see Fig.~\ref{figS2}(b)).
Such a gain or loss changes the transfer matrix of a link waveguide from left to right by a factor of $e^h$, resulting in an effective hopping from $\kappa$ to $\kappa e^h$, where $\kappa$ refers to the hopping when $h=0$. Similarly, the effective hopping from right to left is modified by a factor of $e^{-h}$.

The dynamics of a wave function is governed by
\begin{equation}
i\partial_t|\Psi\rangle=H|\Psi\rangle+\sum_n (\omega_n-\bar{\omega}) \left(\begin{array}{c}
                  a_n|A_{n}\rangle \\
                  b_n|B_{n}\rangle
                \end{array}\right) -i\gamma |\Psi\rangle,
\end{equation}
where
\begin{equation}
|\Psi\rangle=\sum_n \left(\begin{array}{c}
                  a_n|A_{n}\rangle \\
                  b_n|B_{n}\rangle
                \end{array}\right)
\end{equation}
with $a_n$ ($b_n$) denoting the electric field of the light in the A (B) site resonator in the $n$th unit cell,
$\gamma$ is the global decay rate of the field, and $\omega_n=\bar{\omega}+\delta\omega_n$ represents the resonant frequency of
each site resonator. To simulate an external gradient potential, each site resonator is engineered to have a resonant
frequency of $\bar{\omega}+\delta \omega_n$ with $\delta \omega_n=\mathcal{E}n$. One can also realize an external ac electric field
by controlling local refractive index~\cite{Fan2004PRLS,Longi2007PRES}.

\end{widetext}

\end{document}